# Descreening of Field Effect in Electrically Gated Nanopores

*Yang Liu[1*], David E. Huber[2], Vincent Tabard-Cossa[2§], and Robert W. Dutton[1]*

[1]Center for Integrated Systems, Stanford University, Stanford, CA, 94305

[2]Genome Technology Center, Stanford University, Palo Alto, CA 94304

**Abstract:** This modeling work investigates the electrical modulation characteristics of field-effect gated nanopores. Highly nonlinear current modulations are observed in nanopores with non-overlapping electric double layers, including those with pore diameters 100 times the Debye screening length. We attribute this extended field-effect gating to a descreening effect, i.e. the counter-ions do not fully relax to screen the gating potential due to the presence of strong ionic transport.

[*]E-mail: yangliu@gloworm.stanford.edu

[§] Current address: Department of Physics, University of Ottawa, Ottawa, ON, Canada

By analogy with semiconductor field effect transistors (FETs), micro- and nano-scale fluidic "transistors" [1-11] have been extensively studied for electrostatic gating of ionic and molecular transport via surface charges [3,6] or buried gate electrodes [1,2,4,5,7-11]. Their potential applications range from biological sensing [10,12,13] to fuel cells [7] and desalination [14]. In contrast to semiconductors, ionic solutions are essentially zero-bandgap conductors and the field effect is effectively screened by mobile counter-ions. The electric double layer extension is commonly regarded to be limited to $\sim 5\Lambda_D$ [15], where $\Lambda_D$ is the Debye screening length (1 nm for 100 mM ionic concentration). Due to this screening limit, nano-fluidic "transistor" devices are usually designed for the regime of overlapping electric double layers [2,5,6,7,8]. This imposes a stringent constraint on device fabrication and system integration, considering that many important applications involve high ionic strength, e.g. bio-sensing under physiological conditions (150 mM) or desalination of seawater (500 mM) and brackish water (10~500 mM). Meanwhile, the extending of the field effect beyond the screening limit has been observed in perm-selective nano-channels in the presence of strong transport [16,17], including a recent report of current rectification in pores with diameters $\sim 500\Lambda_D$ [18]. Previously, Daiguji et al. simulated the modulation of ionic current by varying surface charge densities in N-P-N bipolar devices with channel diameters $<\sim 10\Lambda_D$ [3]. In their work, the results were interpreted based on a 1-D flux analysis. Nevertheless, for rational design of active nano-fluidic devices, further studies are still needed to elucidate the physical origin of this extended field effect, particularly the role of the inherent coupling between the transversal gating electrostatics and the longitudinal transport of ions and fluids.

In this Letter, we numerically study the modulation characteristics of electrically-gated nanopores with large diameters (~10$\Lambda_D$ and ~100$\Lambda_D$, respectively). In particular, we interpret the observed extended field effect based on a de-screening picture. Onsager et al. previously showed that, under high field strength comparable to $k_BT/q\Lambda_D$, the counter-ion "atmosphere" around a charged particle is less developed due to their relative movement and only forms a partially screening layer [19,20]. The term $k_BT/q$ above represents the thermal voltage. In our earlier work, we studied this descreening effect for long-range biological charge sensing in nanopores [21] and nanowires [22]. Here, we reason that, under strong transport, the same effect also applies to the electric double layers formed at the nanopore gate surfaces, thereby enabling long-range, electrostatic manipulation of charged species. Furthermore, we note that the modulation characteristics studied in this Letter are intrinsically related to electro-kinetics such as limiting and overlimiting conductance, rectification, concentration polarization and vortex formation, which were a subject of our previous work [11].

The device under study (Fig. 1a) is a cylindrically symmetric pore connecting two electrolyte reservoirs separated by a solid-state membrane. The drain bias ($V_d$) between the drain electrode and the grounded source electrode drives the transport. The gate electrode buried inside the oxide dielectrics modulates the transport through gate biasing ($V_g$). We model the ionic transport within the pore and reservoirs using the continuum-based Poisson-Nernst-Planck (PNP) equations:

$$\nabla \cdot (\varepsilon_w \nabla \psi) + q(C_+ - C_-) = 0,$$

$$q\nabla \cdot (-D_+ \nabla C_+ - \mu_+ C_+ \nabla \psi + C_+ \vec{u}) = 0,$$

$$-q\nabla \cdot \left(-D_-\nabla C_- + \mu_- C_-\nabla \psi + C_-\bar{u}\right) = 0,$$

where $\psi$ is the electrostatic potential, $C_\pm$ the ion concentrations, $\varepsilon_w$ the solution permittivity, $\mu_\pm$ the ion mobilities, $D_\pm$ the ion diffusion coefficients, and $\bar{u}$ the solvent velocity. The + and - subscripts indicate the cation and anion species, respectively. The bulk ionic concentration, $C_0$, is approached at the top and bottom boundaries.

We apply the Poisson equation within the oxide and the continuity of $\psi$ across oxide interfaces. The oxide layers are assumed to be impermeable to ions. The gate region is assumed to be equi-potential at $V_g$. To highlight the electrical gating effect, we model a charge-neutral nanopore surface. In practice, surface charges also contribute to the ionic modulation and can be adjusted as an additional design parameter by either pH control or surface chemistry.

We model the fluid transport as an incompressible, Newtonian Stokes flow governed by the Stokes-divergence equations

$$-\nabla p + \gamma \Delta \bar{u} - q(C_+ - C_-)\nabla \psi - k_B T \nabla (C_+ + C_-) = 0,$$

$$\nabla \cdot \bar{u} = 0,$$

where $p$ is the solvent pressure and $\gamma$ the solvent viscosity. Here, the gradient of the excess ionic osmotic pressure is explicitly treated as a body force for improved numerical stability. Boundary conditions for the Stokes equation include: no-slip for the channel surfaces; slip for the symmetry axis; zero pressures and zero normal velocity gradients at the top and bottom reservoir boundaries. The no-slip condition is appropriate for the hydrophilic channel surface assumed in this study, for which the complication of hydrodynamic slippage has been experimentally confirmed to be insignificant [23].

Some physical parameters include: $\varepsilon_w = 80\varepsilon_0$ for water where $\varepsilon_0$ is the vacuum permittivity; symmetric ion mobilities $\mu_+ = \mu_- = 7.62 \times 10^{-8}$ m$^2$/Vs for KCl; $\varepsilon_{ox} = 3.9\varepsilon_0$ for oxide; and $\gamma = 0.001$ Ns/m$^2$ for water. The Einstein relation $D_\pm = \mu_\pm k_B T / q$ is used. More detailed model descriptions, including its numerical validations, are given in the Supplemental Material [24]. For a given set of electrical biases, all of the above transport equations are self-consistently solved over the entire device structure, giving the steady-state, terminal I-V characteristics, $I_d(V_d, V_g)$.

The drain current vs. gate voltage ($I_d$-$V_g$) characteristics are shown for two bulk ion concentrations in Fig. 1b and 1c, respectively. Each $I_d$-$V_g$ curve corresponds to a specific $V_d$ that ranges between 0 V and 3 V. For the 1 mM case in Fig. 1b, $\Lambda_D$ is ~10 nm. It is observed that, under sufficiently high $V_d$'s, $I_d$ is significantly modulated by the gate biasing, even though the pore radius ($R_0$=50 nm) is considerably larger than $\Lambda_D$. Each $I_d$-$V_g$ curve exhibits a symmetry, $I_d(V_d,V_g)=I_d(V_d,V_d-V_g)$, as expected from the symmetries in device geometry and in ion mobilities. The peak current occurs at the symmetric bias condition, $V_d=2V_g$. As $V_g$ shifts away from the symmetric condition in either direction, significant $I_d$ suppression occurs. In the following, we define the gating potential, $\Delta V_g=V_g-V_d/2$, which is accounted from the symmetric condition.

Similar trends in $I_d$-$V_g$'s are observed in Fig. 1c for the 100 mM case, for which $\Lambda_D$ is reduced to ~1 nm. Remarkably, despite the fact that the pore radius is ~50$\Lambda_D$, we still observe an appreciable $I_d$ modulation at sufficiently high $V_d$'s.

For comparison, we consider in Fig. 1d an additional case of fully overlapping electric double layers, where $R_0$=5 nm~$1/2\Lambda_D$ for $C_0$ of 1 mM. In contrast to the previous two cases, $I_d$ monotonically increases as $V_g$ shifts away from the symmetric conditions. This reveals the distinctive difference between the ambipolar-dominant transport in the case of non-overlapping electric double layers and the unipolar-dominant transport in the overlapping case, as previously noted [3,15].

To correlate the observed modulation with the field-effect gating, we study a specific bias condition corresponding to a $\Delta V_g$ of 1 V ($V_d$=2 V and $V_g$=2 V) for the two cases of non-overlapping electric double layers. In Fig. 2a, both $\psi$ and the vertical electric field strength are plotted along the longitudinal axis. Due to the gating effect, $\psi$ drops more rapidly at the bottom side. This leads to regions of strong and weak electric fields at the channel bottom and top portions, respectively. The normalized ion concentration, $\overline{C} = (C_+ + C_-)/2C_0$, is plotted along the longitudinal axis in Fig. 2b. Concentration polarization is clearly observed with the formation of ion depletion and accumulation zones that correspond to the high and low electric field regions, respectively, as a result of ion flux continuity. The presence of concentration polarization, particularly of the ion depletion zone, leads to current limiting behavior [15] and explains the observed current suppression under asymmetric bias conditions. We note that the gating potential is also the underlying cause of electro-osmotic flow that further enhances the current suppression [11].

Central to this Letter is the question as to how the field effect extends from the gate surface to the longitudinal axis, far beyond $\Lambda_D$? In Fig. 3a, we examine the impact of $V_d$ on the electrostatic potential change, $\Delta\psi$, induced by a fixed gating potential, $\Delta V_g$=1 V. $C_0$ of 1 mM is used in this example. $\Delta\psi$ is obtained as the difference between the potential profile with the gating potential applied and that without (i.e. the symmetric condition). For the thermal equilibrium condition ($V_d$=0 V), $\Delta\psi$ is fully screened and decays exponentially from the gate surface, in agreement with the common electric double layer model. Such an exponential decay is a direct result of the detailed balance between the drift and diffusion processes of mobile ions, which is no longer satisfied in the presence of ionic flux [21]. We clearly observe that, for $V_d$=1 V, the gating potential is only partially screened and has a significant portion extending toward the longitudinal axis. Such de-screening is even more evident for $V_d$=2 V.

We further examine the cation-anion concentration difference, $D = C_+ - C_-$, a quantity directly proportional to the net charge density. In Fig. 3b, we plot the profiles of $\Delta D$, which is the change in $D$ induced by a fixed $\Delta V_g$ of 1 V, for various $V_d$'s. As $V_d$ increases from 0 V to 2 V, the magnitude of $\Delta D$ decreases dramatically, thus becoming less effective at shielding the gating potential. This is a further evidence of the descreening effect.

To quantify the descreening effect, we specifically examine the $V_d$ dependence of $\Delta\psi_c$, the potential change at the device center point due to the fixed $\Delta V_g$ of 1 V, in Fig. 3c. Results for both $C_0$ values, 1 mM and 100 mM, are shown. In the same figure, we also plot the $V_d$ dependence of $|\Delta D_m|$, the magnitude of $\Delta D$ at the middle point of the gate surface. Correlation between the two quantities is consistently observed. As the level of ion transport increases with $V_d$, the amount of induced screening charge proportional to $|\Delta D_m|$ is significantly reduced. Correspondingly, as a result of this descreening effect, $\Delta\psi_c$ reaches 0.76 V and 0.22 V for 1 mM and 100 mM $C_0$'s, respectively, under a $V_d$ bias of 3 V.

In summary, we have numerically investigated the modulation characteristics in electrically gated nanopores with non-overlapping electric double layers. It is revealed that the field effect is extended far beyond the Debye screening length and results in nonlinear current modulation, which is appreciable even in nanopores with diameters ~100$\Lambda_D$. We attribute such an extended field effect to the descreening of counter-ions at the gate surfaces under strong ion transport.

**ACKNOWLEDGMENTS:** We appreciate discussions with R. Howe, J. Sauer, J. Snapp, J. Santiago, T. Zangle, and C. Rafferty.

References:
x

x

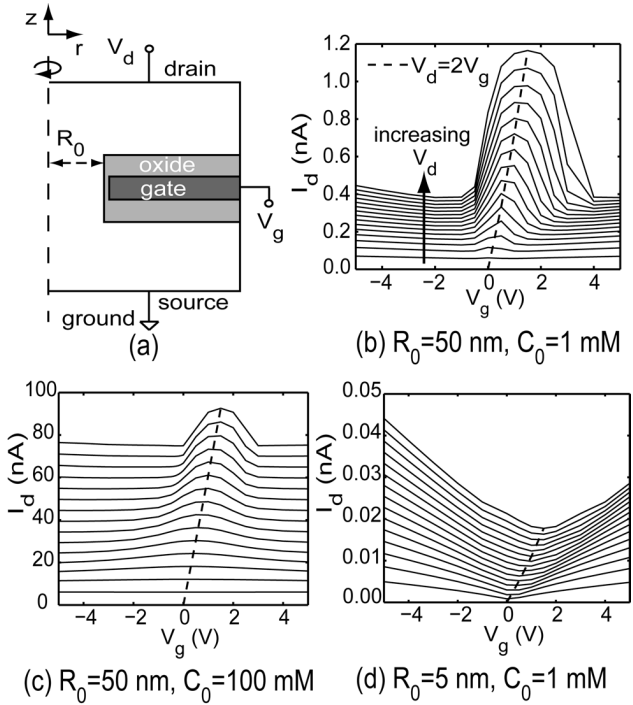

Fig. 1: (a) Schematic of a gated nanopore device with cylindrical symmetry (not to scale). Some device parameters include: top and bottom oxide thickness 100 nm each; gate electrode thickness 100 nm; side-wall gate oxide thickness 2 nm; reservoir size 1 μm in both width and thickness; (b)(c)(d) $I_d$ vs. $V_g$ characteristics for constant $V_d$ values that range from 0 V to 3 V at a step of 0.2 V. The dashed curve corresponds to the current at symmetric bias conditions, $I_d(V_d=2V_g, V_g)$. The pore radius ($R_0$) and bulk ion concentration ($C_0$) values are specified for each case.

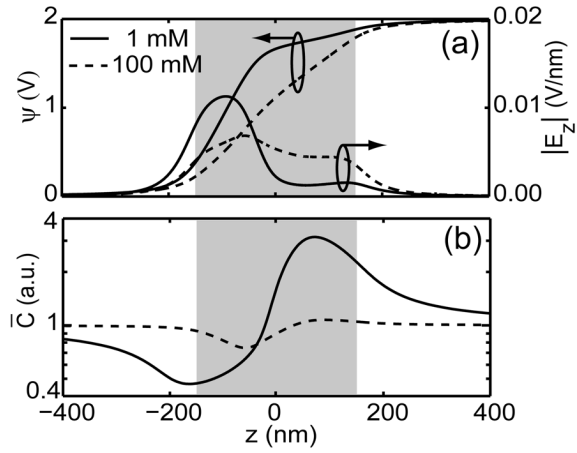

Fig. 2: (a) Simulated profiles of electrostatic potential and vertical electric field strength along the longitudinal axis for a bias condition $V_d = V_g = 2$ V under two $C_0$ conditions, 1 mM and 100 mM; (b) profiles of normalized ion concentration, $\overline{C} = (C_+ + C_-)/2C_0$, along the longitudinal axis for these two cases. Only the portion of interest is shown and the shaded areas indicate the nanopore region.

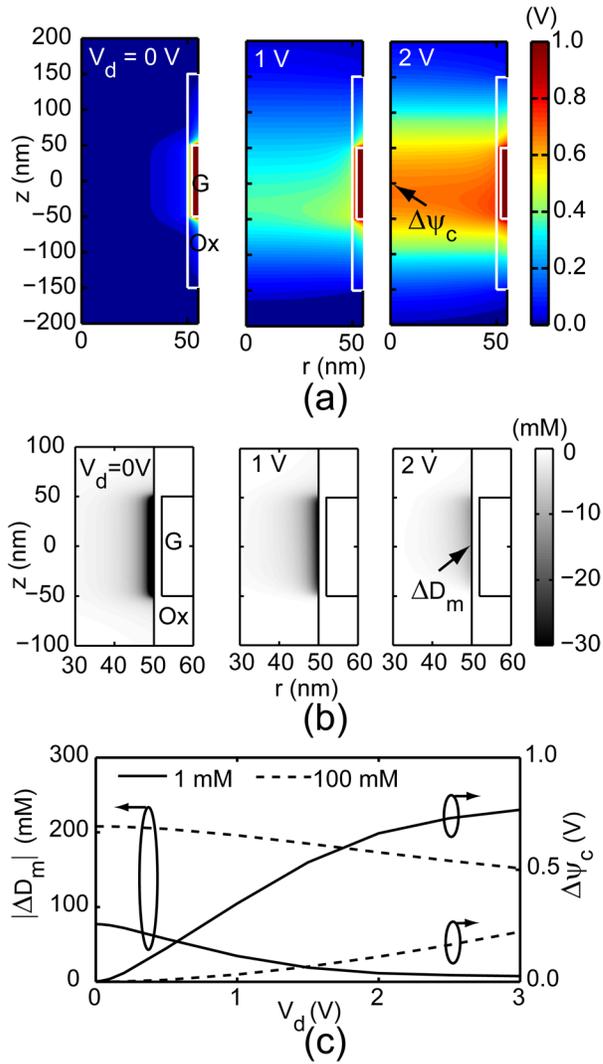

Fig. 3: Profiles of (a) $\Delta\psi$ and (b) $\Delta D$ that are induced by a fixed $\Delta V_g$ of 1 V under three $V_d$ biases. Only the portions of interest are shown; (c) dependence on $V_d$ of both $\Delta\psi$ at the device center point ($\Delta\psi_c$) and the magnitude of $\Delta D$ at the middle point of the gate surface ($|\Delta D_m|$).